\documentstyle[editedvolume,numreferences]{crckapb}

\newcommand{\be}{\begin{equation}}
\newcommand{\ee}{\end{equation}}

\def\unlockat{\catcode`\@=11}

\def\lockat{\catcode`\@=12}

\unlockat

\def\newsec#1{\global\advance\secno by1\message{(\the\secno. #1)}
\global\subsecno=0\global\subsubsecno=0\eqnres@t\noindent
{\bf\the\secno. #1}
\writetoca{{\secsym} {#1}}\par\nobreak\medskip\nobreak}

\global\newcount\subsecno \global\subsecno=0
\def\subsec#1{\global\advance\subsecno by1\message{(\secsym\the\subsecno. #1)}
\ifnum\lastpenalty>9000\else\bigbreak\fi\global\subsubsecno=0
\noindent{\it\secsym\the\subsecno. #1}
\writetoca{\string\quad {\secsym\the\subsecno.} {#1}}
\par\nobreak\medskip\nobreak}

\global\newcount\subsubsecno \global\subsubsecno=0
\def\subsubsec#1{\global\advance\subsubsecno by1
\message{(\secsym\the\subsecno.\the\subsubsecno. #1)}
\ifnum\lastpenalty>9000\else\bigbreak\fi
\noindent\quad{\secsym\the\subsecno.\the\subsubsecno.}{#1}
\writetoca{\string\qquad{\secsym\the\subsecno.\the\subsubsecno.}{#1}}
\par\nobreak\medskip\nobreak}

\def\subsubseclab#1{\DefWarn#1\xdef #1{\noexpand\hyperref{}{subsubsection}%
{\secsym\the\subsecno.\the\subsubsecno}%
{\secsym\the\subsecno.\the\subsubsecno}}%
\writedef{#1\leftbracket#1}\wrlabeL{#1=#1}}
\lockat

\def\IB{\relax\hbox{$\inbar\kern-.3em{\rm B}$}}
\def\IC{\relax\hbox{$\inbar\kern-.3em{\rm C}$}}
\def\ID{\relax\hbox{$\inbar\kern-.3em{\rm D}$}}
\def\IE{\relax\hbox{$\inbar\kern-.3em{\rm E}$}}
\def\IF{\relax\hbox{$\inbar\kern-.3em{\rm F}$}}
\def\IG{\relax\hbox{$\inbar\kern-.3em{\rm G}$}}
\def\IGa{\relax\hbox{${\rm I}\kern-.18em\Gamma$}}
\def\IH{\relax{\rm I\kern-.18em H}}
\def\II{\relax{\rm I\kern-.18em I}}
\def\IK{\relax{\rm I\kern-.18em K}}
\def\IL{\relax{\rm I\kern-.18em L}}
\def\IP{\relax{\rm I\kern-.18em P}}
\def\IQ{\relax{\rm I\kern-.42em Q}}
\def\IR{\relax{\rm I\kern-.18em R}}
\def\IS{\relax{\rm I\kern-.18em S}}

\def\IZ{\relax\ifmmode\mathchoice{\hbox{\cmss
Z\kern-.4em Z}}{\hbox{\cmss Z\kern-.4em Z}}
{\lower.9pt\hbox{\cmsss Z\kern-.4em Z}}
{\lower1.2pt\hbox{\cmsss Z\kern-.4em Z}}\else{\cmss Z\kern-.4em Z}\fi}

\def\CA {{\cal A}}
\def\CB {{\cal B}}
\def\CC {{\cal C}}
\def\CD {{\cal D}}

\def\CF {{\cal F}}

\def\CH {{\cal H}}

\def\CL {{\cal L}}
\def\CM {{\cal M}}
\def\CN {{\cal N}}
\def\CO {{\cal O}}
\def\CP {{\cal P}}
\def\CQ {{\cal Q}}

\def\CU {{\cal U}}

\def\CW {{\cal W}}

\def\p{\partial}

\def\Det{{\rm Det}}

\def\p{{\partial}}

\font\manual=manfnt \def\dbend{\lower3.5pt\hbox{\manual\char127}}

\def\half {{1\over 2}}

\def\Det{{\rm Det}}

\def\Imm{{\Im}{\rm m}}
\def\Ree{{\Re}{\rm e}}
\def\inbar{\,\vrule height1.5ex width.4pt depth0pt}

\def\lieg{{\underline{\bf g}}}

\def\liet{{\underline{\bf t}}}

\font\cmss=cmss10 \font\cmsss=cmss10 at 7pt

\def\boxit#1{\vbox{\hrule\hbox{\vrule\kern8pt
\vbox{\hbox{\kern8pt}\hbox{\vbox{#1}}\hbox{\kern8pt}}
\kern8pt\vrule}\hrule}}

\def\mathboxit#1{\vbox{\hrule\hbox{\vrule\kern8pt\vbox{\kern8pt
\hbox{$\displaystyle #1$}\kern8pt}\kern8pt\vrule}\hrule}}

\def\inbar{\,\vrule height1.5ex width.4pt depth0pt}

\font\cmss=cmss10 \font\cmsss=cmss10 at 7pt

\begin{opening}
\title{Testing Seiberg-Witten Solution}
\author{A.~ Losev}
\institute{Institute of Theoretical and Experimental Physics,\\
117259, Moscow, Russia}
\author{ N.~ Nekrasov}
\institute{Institute of Theoretical and Experimental Physics,\\
117259, Moscow, Russia\\
 Lyman Laboratory of Physics, Harvard University,\\
Cambridge MA 02138}
\author{S.~ Shatashvili}
\institute{Department of Physics, Yale University, \\New Haven  CT
06520, Box 208120\\
on leave of absence from \\
St. Petersburg Steklov Mathematical Institute}
\subtitle{HUTP-97/A102, ITEP-TH-74/97, hep-th/9801061\\
To appear in the Proceedings of Cargese conference, June 1997}
\end{opening}
\runningtitle{Testing Seiberg-Witten Solution}

\begin{document}
\begin{abstract}
We propose a few tests of Seiberg-Witten solutions
of $\CN=2$ supersymmetric gauge theories
by the  instanton calculus in  twisted gauge theories.
We re-examine
the low-energy effective abelian theory in the presence of
sources and  present the
formalism which makes duality transformations transparent
and easily fixes all the contact terms in a broad class of theories.
We also discuss ADHM integration and its relevance to the stated problems.

\end{abstract}

\section{Introduction and summary}

One of the
sources of the recent duality revolution in string theory is the
solution of N.~Seiberg and E.~Witten \cite{SeWi} of $\CN=2$ supersymmetric
Yang-Mills theory which has been tested in many indirect ways
but never directly. The solution, among other things,
predicts the formula for the effective coupling constant of
the low-energy theory as a function of the order parameter $u$
(in $SU(2)$ theory):
\be
\label{swcl}
{{\tau}(u) = {{2i}\over{\pi}}
{\rm log}\left({u\over{\Lambda^{2}}}\right) + \sum_{n=1}^{\infty}
{\tau}_{n} \left( {{\Lambda^{2}}\over{u}} \right)^{2n}}\ee
The coefficients $\tau_{n}$ are claimed to be instanton corrections.
The direct test of (\ref{swcl}) would involve integration
of a certain form over the moduli space of instantons on $\IR^{4}$.
This integration can be shown to localize onto the space
of {\it point-like instantons}
  and has a potential divergence in it
related to the fact that the space of point-like
instantons is non-compact (it is a resolution of singularities
of $S^{k}{\IR}^{4}$). Although this difficulty is
avoidable by appropriate regularization
so far no substantial success has been achieved on this route
(nevertheless, see the
related discussion in \cite{khoze}). The infrared regularization
of the instanton integrals turns out to be tricky problem. At first sight
it is possible to simply put the periodic
boundary conditions and study the theory on a four-torus.
Unfortunately the puriest test of (\ref{swcl}) would involve
the computation of the correlation functions of observables which preserve
some supersymmetry which are called topological correlation functions\index{correlation functions of the topological observables}. TCF's  are almost insensitive to the geometry
of the moduli space of vacua of the gauge theory in the case, where
the four-manifold $X$ has $b_{2}^{+}>1$ . To make the TCF probe the geometry of
the moduli space of vacua one must find such a manifold $X$ which  allows
for the instanton calculus to be perfomed and has $b_{2}^{+}=1$. The latter constraint will be explained below. Examples of such manifold $X$ are provided by ruled surfaces,
 $X = S^{2} \times S^{2}$ among them .  The manifold $X$ does not have any
covariantly constant spinors and the original supersymmetric theory has
no conserved
charges. One may study  a twisted theory, introduced by E.~Witten in  \cite{witttft}.
The latter has more
freedom in the choice of action then the original physical
theory. In particular, one may take the limit
of zero coupling in the non-abelian theory which reduces
the computations of TFC to the counting intersection numbers of
homology classes of compactified moduli space of instantons ion $X$.
Such a counting  problem is equivalent
 to the way S.~Donaldson formulated his
invariants in the language of problems of ``generic position''
(for the gauge groups $SU(2)$ and $SO(3)$).
The insight coming from the equivalence of Donaldson
theory and Witten's twisted version of $\CN=2$ super-Yang-Mills
theory is that the same problem may be addressed in
the infrared limit.
It seems that the well-defined ultraviolet problem
in the field-theoretic formulation involves
the definition of the compactified moduli space of {\it distinct}
points on the space-time manifold $X$.  The infrared
theory contains the information about this compactification
in the contact terms between the observables
\index{contact terms}. In the context
of gauge theory these contact terms can be studied
using severe constraints of modular invariance and ghost number
anomalies
\index{modular invariance, ghost number anomalies}.

\section{The micro/macroscopic theories}

The $\CN=2$ super-Yang-Mills theory with the gauge group $G$
has as the basic field the vector multipler of four dimensional
$\CN=2$ supersymmetry, which involves a gauge field $A^{a}$, a complex
scalar $\phi^{a}$ and a pair of Weyl fermions $\psi_{\alpha}^{a},
\lambda^{\dot\alpha, a}$. All fields are in the adjoint representation of
the gauge group.
For the sake of convenience we study the twisted
version of the theory where fermions have different
Lorentz spins. One should understand once and for all
that as long as ``physical'' questions are asked about
the theory formulated on $\IR^{4}$ (euclidean signature) there is
no way to distinguish the two models. The distinction appears
when the curved gravitational background appears. The twisted theory always has at least one conserved supercharge $Q$.
The square of $Q$ is an infinitesimal
 gauge transformation with parameter $\phi$.
The operator $Q$ acts on the Hilbert space of the theory. Its square
vanishes on the physical Hilbert space formed by the gauge invariant
states.
The Hilbert space $\CH_{S}$ is obtaned by quantizing the space of
fields on a three-fold $S$.
 The Hamiltonian of the theory can be represented as
a $Q$-commutator:
\be
\label{hmltn}
H = \{ Q, G_{r} \}
\ee
where $G_{r}$ is a certain operator which is in fact a twisted
version of one of the eight  supercharges of the original $\CN =2$ susy.
There is an operator of ghost number $U$ which can be represented
as the integral of
the corresponding current $J$: $U = \int_{S} J$.
The Hamiltonian has ghost charge $0$, the operator $Q$ has $U=1$,
$G_{r}$ has $U=-1$, the fields have the following charges:
\begin{table}[chrgs]
\begin{center}
\caption{Ghost charges, Lorentz spins, $Q$ action}
\begin{tabular}{lllll}
\hline
{\rm Field} & $SU(2)_{L} \times SU(2)_{R}$ & $U$ {\rm charge}& {\rm
Statistics} & $Q$ {\rm action} \\
\hline
$A$ & \quad $(\half, \half )$ & $0$ & B & $\psi$ \\
$\psi$ & \quad $(\half, \half)$ & 1 & F & $D_{A}\phi$ \\
$\phi$ & \quad $(0,0)$ & 2 & B & $0$ \\
$\chi$ & \quad $(0, 1)$ & -1 & F & $H$\\
$H$ & \quad $(0,1)$ & 0 & B & $[\phi, \chi ]$\\
$\eta$ & \quad $(0,0)$ & -1 & F & $[\phi, \bar\phi]$ \\
$\bar\phi$ & \quad $(0,0)$ & -2 & B & $\eta$\\
\hline
\end{tabular}
\end{center}
\end{table}

Since on the gauge invariant states
$Q^{2}=0$ one may study the cohomology space $H_{S} = {\rm Ker}Q/
{\rm Im}Q$. The importance of this space is that it is preserved under the
evolution and changes of the metric. Indeed, if $Q \vert \psi_{0} \rangle =0$ and
we identify the states in the evolved space with the help
of Hamiltonian $H$ then
\be
\label{evol}
\vert \psi_{t} \rangle = e^{ - t H} \vert \psi_{0} \rangle =
\vert \psi_{0} \rangle  + Q \vert \psi^{\prime}_{t} \rangle
\ee
that is the cohomology class of a state does not change with time.
Now, having got the space $H_{S}$ one may wonder
how many states are there and what are the natural operators which act
on $H_{S}$. To answer the second question we must figure out
what are the operators $\CO$ the  theory which obey
$\{ Q, \CO \} = 0$. The Hamiltonian is one of them but it acts
trivially on $H_{S}$. What are non-trivial operators? Clearly,
these are the ones which cannot be represented as
$\{ Q, {\rm smth} \}$, that is we are interested in $Q$-cohomology
in operators.

The twisted gauge theory has various non-local observables. They are labelled by
the following data: a gauge  invariant  function $\CP$ on the Lie algebra $\lieg$ and the
number $p= 0, \ldots, 4$. The $p$-observable
$\CO^{(p)}_{\CP}$ is a $p$-form and is supposed to be  integrated over a $p$-cycle in the manifold $X$.  The easiest way to write all $p$-observables at once is to
expand the function $\CP(\phi + \psi + F)$ is the forms of varios degree. The observables
$\CO^{(p)}$ then automatically obey the descend equations
\be
\label{dscnd}
d \CO^{(p)} = \{ Q, \CO^{(p+1)} \}
\ee
Of course, the representative of $Q$-cohomology is not unique. The one we get by
expanding  $\CP(\phi + \psi + F)$ is called {\it holomorphic}. For some reasons
one may prefer a harmonic representative, which is obtained
from $\CP(\phi)$ by acting with $G_{\mu} $ on it. Here $G_{\mu}$ are the twisted
supercharges, which obey $\{ Q, G \} = d$ (we saw one of them in (\ref{hmltn})).
The  holomorphic observables can be written quite explicitly:
$\CO^{(0)} _{\CP}  =  {\CP} ({\phi})$,  $\CO^{(1)}_{\CP} =
{{\p {\CP}}\over{\p \phi^{a}}} \psi^{a}$, $\CO^{(2)}_{\CP} =
{{\p {\CP}}\over{\p \phi^{a}}} F^{a}  + {\half} {{\p^{2}
{\CP}}\over{\p \phi^{a} \p \phi^{b}}} \psi^{a}  \psi^{b} $, $\CO^{(3)}_{\CP} =  {\half} {{\p^{2} {\CP}}\over{\p \phi^{a} \p \phi^{b}}} \psi^{a}  F^{b}  + {1\over{6}} {{\p^{3}
{\CP}}\over{\p \phi^{a} \p \phi^{b} \p \phi^{c}}} \psi^{a}  \psi^{b} \psi^{c} $.
The top degree observable equals:
\be\CO^{(4)}_{\CP} =
{\half} {{\p^{2} \CP}\over{\p \phi^{a} \p \phi^{b}}} F^{a}  F^{b} +
{1\over{3!}} {{\p^{3}\CP}\over{\p \phi^{a} \p \phi^{b} \p \phi^{c} }}
F^{a} \psi^{b}  \psi^{c} + {1\over{4!}} {{\p^{4} \CP}\over{\p \phi^{a}
\p\phi^{b} \p \phi^{c} \p \phi^{d}}} \psi^{a} \psi^{b} \psi^{c} \psi^{d}
\label{top}\ee
It enters the Seiberg-Witten low-energy effective
action, where all the fields are specialized to be abelian. In general, the
whole action $S$ equals the sum of the
$4$-observable, constructed
out of the prepotential $\CF$ and the $Q$-exact term:
\be\label{actn}{S = \int_{X} \CO^{(4)}_{\CF} + \{ Q, R \} . } \ee
We can view all observables $\CO^{(p)}_{\CP}$ for all $\CP$ and $p$
as deformations of the theory by adding them to the action (\ref{actn}) as follows:
\be\label{actni}
S = \sum_{p=0}^{4} T^{p, \alpha}
\int_{X} \left( \CO^{(p)}_{\CF_{p}}  + \{ Q, R^{(p)} \} \right) \wedge e_{\alpha}
\ee
where $e_{\alpha}$ runs through a basis in the cohomology group $H^{*}(X; \IR)$.
In turn, one may expand $T^{p, \alpha} \CF_{p} = \sum_{k} T^{k, \alpha} \CP_{k} $ with $\CP_{k}$
running over a space of invariant polynomials.

The infrared theory is also
a twisted supersymmetric gauge theory with abelian gauge group. The observable
$\CO^{(p)}_{\CP, \rm UV}$ of the ultraviolet theory flows to the
observable $\CO^{(p)}_{\CP, \rm IR}$ of the infrared theory. Since the gauge group of the infrared theory is the maximal torus (extended by the Weyl group) of the ultraviolet theory there is one-to-one correspondence
between the gauge invariant functions $\CP_{\rm UV}$ and $\CP_{\rm IR}$.
Now suppose that two cycles $C^{p} \in H_{p}(X)$ and $C^{q} \in H_{q}(X)$
intersect: $C_{p} \cap C_{q} = C_{p+q-4} \in H_{p+q-4}(X)$. No matter
how large the scale is to get at the intersection point(s) we must look at very short distances.
The contribution of the short distance physics is the contact term:

\be
\label{ct}
\int_{C^{p}} \CO^{(p)}_{\CP, \rm UV} \int_{C^{q}} \CO^{(q)}_{\CQ, \rm UV} \longrightarrow
\int_{C^{p}} \CO^{(p)}_{\CP, \rm IR} \int_{C^{q}} \CO^{(q)}_{\CQ, \rm IR}  +
\int_{C^{p+q-4}} \CO^{(p+q-4)}_{\CC(\CP, \CQ), \rm IR}
\ee

The main problem of the effective usage of low-energy theory is the determination
of the contact terms $\CC(\CP, \CQ)$. The important constraint on the
function $\CC(\CP, \CQ)$
is that it is independent of $p$ and $q$. This property was called in
 \cite{issues} {\it
the universality of the contact terms}. It follows from the simple
fact that the {\it relative} geometry
of a pair of points on a four-fold $X$ is independent of the dimensions
of the cycles
the points are confined to.  This universality property implies some
 non-trivial idenities
between the modular functions which we describe in the next section.
These identities
seem to distinguish Seiberg-Witten-like families of abelian varieties
and serve as a test of SW technology.

The next contact term is the triple  term : suppose three cycles
$C^{p_{1}}, C^{p_{2}}, C^{p_{3}}$
are given, then the analogue of (\ref{ct}) is:
\begin{eqnarray}
& \prod_{i=1}^{3} \int_{C^{p_{i}}} \CO^{(p_{i})}_{\CP_{i}, \rm UV} \longrightarrow
\prod_{i=1}^{3} \int_{C^{p_{i}}} \CO^{(p_{i})}_{\CP_{i}, \rm IR}  + \nonumber\\
&  \sum_{(i,j,k) = (1,2,3)}
\int_{C^{p_{i}} \cap C^{p_{j}}} \CO^{(p_{i}+p_{j}-4)}_{\CC(\CP_{i},
\CP_{j}), \rm IR} \int_{C^{p_{k}}}
\CO^{p_{k}}_{\CP_{k}, \rm IR} \nonumber\\
&  +
\int_{C^{p_{1}} \cap C^{p_{2}} \cap C^{p_{3}} } \CO^{p_{1}+p_{2}+
p_{3}-4}_{\CC(\CP_{1}, \CP_{2}, \CP_{3}), \rm IR} \label{ctt}
\end{eqnarray}

If the four-observables are involved then
we get an infinite sequence of contact functions
$\CC(\CP_{1}, \ldots, \CP_{k})$ which are nicely
organized in the generating function $\CF$:
\be
\label{cntfnc}
\CF (T) = \CF_{0} + \sum_{k=1}^{\infty}  T^{1} \ldots T^{k} \CC(\CP_{1}, \ldots, \CP_{k})
\ee
where $T^{k}$'s are the deformation parameters, which wil be
referred to as  times.
The first order term in the function $\CF$ is $T^{k}\CC(\CP_{k})$.
The ``one-contact function''
$\CC(\CP_{k})$ is the function $\CP_{k}$ itself.

The claim is finally:

{\it The TCF (= partition function) in the theory
with the action (\ref{actni}) is
equal to that in the low-energy abelian theory with the action}
\be
\label{actnii}
S = \int_{X}  \CF(a + \psi + F, T^{k, \alpha}e_{\alpha})
\ee
In \cite{issues} it is shown  that the function
$\CF(T)$ has a very simple geometrical meaning. It describes the deformations
of $\Gamma$-invaraint Lagrangian submanifolds
in a complex vector symplectic space $\IC^{2r}$,
where $\Gamma$ is a certain
discrete subgroup of linear symplectic group.

Similarly to the studies of two dimensional type $\bf B$
topological sigma models a lot of information can be rather easily
obtained by working with holomorphic representatives. This is the point
of view taken in this lecture. The reader interested in harmonic theory should
consult \cite{grwt}, \cite{issues}, \cite{grm}.

\section{Coulomb branch measure}

The general strategy in using the statement (\ref{actnii}) is to compute
the integral over the zero modes of the fields entering abelian multiplets.
The final integral is the one over the scalar field $a$ and it involves
various modular forms integrated over the modular domain. Without
going into tremendous complications of the actual comutations
in any realistic setting one may say a lot of non-trivial
things about the measure itself. To compare the last sentence with the
development of perturbative string theory one may
say that the measure of stringy loop computations has been
stduied extensively although the actual integral is known only
in a limited number of instances.

First of all we must explain what distinguishes the manifolds with
$b_{2}^{+}=1$. This issue is not possible to see working with holomorphic
fields only. It turns out that the harmonic representative of
the four-observable contains the only one term ( $\int_{X} \eta \chi \wedge
F$) which couples to a zero mode of  $\eta$ and survives in the
limit of very large $X$. Thus one needs presicely one zero mode
of the field $\chi$ otherwise the effective measure vanishes as
an inverse power of the size of manifold $X$.

It turns out that the TCF's are not exactly
topological invariants. The reason is  that sometimes the enumerative
problem of Donaldson's is not well-behaved under the variations
of parameters, such as the metric. This occurs precisely
when $b_{2}^{+} \leq 1$ due to the abelian instantons\footnote{They are
honest solutions of instanton equations with abelian gauge group}.
The jumps of the correlations functions/Donaldson invariants for $b_{2}^{+}=1$
\index{Donaldson invariants}
are under control \cite{gottshi}, \cite{gz},\cite{borch},\cite{grwt}.

Below we describe a few techinques of computing the pair contact terms
in various theories with   simply-laced gauge groups\footnote{The contact term
between the two-observables constructed out of
quadratic casimirs and certain steps in extending
the integrals of \cite{grwt} to the case of $SU(N)$ gauge group are
independently obtained in \cite{grm}}.
 We use  the low-energy effective theory, whose  action
on $\IR^{4}$
has been computed in \cite{SeWi}, and certain
aspects of it for the general four-manifold $X$
have been worked out in \cite{wittabl} and also recently in \cite{grwt},
\cite{issues}, \cite{grm}.

\subsection{Low-energy theory}
 The low-energy theory contains $r$
$\CN=2$ vector multiplets, which
are defined up to $\Gamma$- transformation, where
$\Gamma$ is a subgroup of $Sp_{2r}({\IZ})$, e.g ${\Gamma}(2)$ or
${\Gamma}_{0}(4)$ for $r=1$.
Let us denote the scalar components of the multiplet, which are monodromy
invariant (up to a sign) at $u^{k} = \infty$ by $a^{i}$.
Then the $S$-dual ones will be denoted as $a_{i,D}$.
The low-energy effective couplings are denoted as
$\tau_{ij} (a) = \left( {{4\pi i }\over{g_{eff}^{2}}} +
{{\theta_{eff}}\over{2\pi}}\right)_{ij} =
 {{\p a_{i,D}}\over{\p a^{j}}}
= \tau_{ij, 1} +i {\tau}_{ij, 2} \equiv \Ree\tau_{ij} + i \Imm\tau_{ij}$

The $Q$- transformations of the gauge fields  must be
consistent with the electric-magnetic duality. It is not clear a priori
that such a $Q$-action  exists, since the duality is a non-local
operation on the fields, while the $Q$ is a local one.
One may try to imitate
the twisted version of the supersymmetric duality transformation
presented in \cite{SeWi}.

We take the approach where only
holomorphuic representatives of the fields are used. It
allows  for a quick check of the modular invariance of the
measure in a relatively simple
setting. One has to  work with (formal)
contour integrals.  Most of the constructions can be done working
with the ``holomorphic'' fields $a, \psi, A$ only. The only
trouble with such a prescription is the absence of Laplacians
and non-definiteness of the topological terms like  $F \wedge F$.
The first problem is avoided in  certain cases by working
with cohomology (with harmonic forms) while the
second may be treated via analytic continuation. In any case
such an approach is useful in getting the right structures.
Once it is done one may introduce the quartet
of the fields ${\bar a}, \eta, \chi, H$ and justify the
constructions by working with the standard
positive-definite actions.

Consider the {\it short superfield}:
\be\label{spfld}{\CA_{i,D} = a_{i,D} + \psi_{i,D} + F_{i,D}}\ee
where $dF_{i,D} = 0$.
The operator $Q$ acts as follows: $Q a_{i,D} = 0$,$Q \psi_{i,D} =
da_{i,D}$,
$QF_{D} = d\psi_{i,D}$.
We impose (by hands) the condition that $F_{i,D}$ represents the
integral cohomology class of the space-time manifold $\Sigma$.
Thus,
$\CA_{i,D} \in \Omega^{0}(\Sigma)_{B} \oplus \Omega^{1}(\Sigma)_{F} \oplus
\Omega^{2}_{\IZ} ({\Sigma})_{B}$
Here $\Omega^{2}({\Sigma})_{\IZ}$ is the space of closed
two-forms with periods in $2\pi i\IZ$.
The indices $B, F$ denote the bosonic and fermionic fields respectively.
The superfield $\CA_{D,i}$ obeys
the condition $(Q-d)\CA_{D,i} = 0$.
One may
also fulfill the condition of $Q-d$-closedeness by
introducing a complete
set of $p$-forms which we call {\it the long superfield}:
\begin{table}[hh]
\begin{center}
\caption{Long superfield}
\begin{tabular}{llllll}
\hline
{\rm Degree of a form} & 0 \quad & 1 \quad & 2 \quad & 3 \quad & 4 \quad\\
\hline
{\rm Field} $\CA^{i}  = $ & $a^{i}$  + &$\psi^{i}$ + &$F^{i}$ + &$\rho^{i}$
+
&$D^{i}$\\
\hline
\end{tabular}
\end{center}
\end{table}
$\CA^{i} = \sum_{p=0}^{4} \CA^{i,p} \in V = \oplus_{p=0}^{4} \Omega^{p}({X})$
and $Q$ acts as $Q\CA^{i,p} = d\CA^{i,p-1}$.
Let $\CF_{D}$ be a holomorphic function on $\IC^{r}$.
The ``action''
\be\label{aci}{S = \int_{\Sigma}\CF_{D} (\CA_{D})}\ee is clearly
$Q$-invariant.
The long superfield $\CA$ allows reparameterizations:
\be\label{dffm}{\CA^{i} \mapsto {\widetilde{\CA^{i}}} (\CA)}\ee
induced by the holomorphic maps $a^{i} \mapsto {\tilde a}^{i}(a^{k})$.
Let $\CL \subset \IC^{2r}$ be a $\Gamma$-invariant
Lagrangian subvariety.
Let $u^{k}, k=1, \ldots r$ be the generators of
the ring $\CW_{\CL}$ of  globally defined $\Gamma$-invariant
holomorphic functions on $\CL$. Extend them to the long superfields
$\CU^{k}, k=1, \ldots, r$. Define the measure
\be\label{msr}{[\CD\CU] = \prod_{k=1}^{r} du^{k} d\psi_{u}^{k} dF^{k}_{u}
d\rho^{k}_{u}dD^{k}_{u}}\ee
where $(Q - d) \left( {u}^{k} + \psi^{k}_{u} + F^{k}_{u} + \rho^{k}_{u} +
D^{k}_{u} \right) = 0$.
The duality transformation proceeds as follows:
introduce both $\CA_{i,D}$ and $\CA^{i}$ and consider the action
$S^{\prime} = \int_{\Sigma}
{\CA}^{i}{\CA}_{i,D} - \CF({\CA}^{i})$.
Let us consider the following (formal) path integral:
\be\label{frmlpi}{
\int \CD \CU^{i} \CD \CA_{i,D} e^{-S}.}\ee
The measure $\CD\CA_{i, D}$
is defined canonically. The dependence of the
measure on $\CA^{i}$ on the choice of the measure
$du^{1} \wedge \ldots du^{r}$ is
completely parallel to  anomaly in Type $\bf B$ sigma models
in two dimensions \cite{ttstar},\cite{Witr}.
The integral over $\CA_{i,D}$ (together with summation over
the fluxes of $F_{i,D}$) forces $D^{i}, \rho^{i}$ to vanish,
while $F^{i}$ becomes a curvature of a connection $A^{i}$.
As a result one gets a measure
$$
{\Det}_{ij} \left( {{\p u^{i}}\over{\p a^{j}}} \right)^{\chi \over{2}}
\prod_{k=1}^{r} d a^{k} d\psi^{k}  dF^{k}
$$
On the other hand, performing
the integral over $\CU$ gives us:
$a_{i,D}  = {{\p \CF}\over{\p a^{i}}}$, $\psi^{i}  =
\left({\tau}^{-1}\right)^{ij}\psi_{j,D}$, $F^{i}
=  \left({\tau}^{-1}\right)^{ij}  \left(
F_{j, D}   - \half ({\tau}^{-1})^{lm} ({\tau}^{-1})^{kp}
(\p^{3}_{lkj}\CF) \psi_{m,D}\psi_{p,D}
\right)$ with $\tau_{ij} = {{\p^{2} \CF}\over{\p a^{i}\p a^{j}}}$.
The determinants in this case are slightly  involved:
\begin{eqnarray}
& {\Det}_{ij} \left( {{\p u^{i}}\over{\p a^{j}}} \right)^{\chi \over{2}}
\left( {\rm Det}\tau\right)^{-{\rm dim}\Omega^{0} + {\rm dim}\Omega^{1}  - {\half}
{\rm dim}\Omega^{2}}  = \nonumber \\
& ({\Det}\tau)^{-{{\chi}\over{2}}}
 {\Det}_{ij} \left( {{\p u^{i}}\over{\p a^{j}}} \right)^{\chi \over{2}} =
\nonumber \\
& {\Det}_{ij} \left( {{\p u^{i}}\over{\p a_{D,j}}} \right)^{\chi \over{2}}
\label{mdlr}
\end{eqnarray}
Were there no
factor ${\Det}_{ij} \left( {{\p u^{i}}\over{\p a^{j}}} \right)^{\chi \over{2}}$
the duality transformation would be anomalous.
This anomaly was already observed in
\cite{wittabl} (for $r=1$).
The ``action'' $S^{\prime}$ evaluates to
(\ref{aci}) with the substitution of $\CF$ by $\CF_{D}$,
which is the Legendre transform of $\CF$.

As it has been explained in \cite{wittabl} the low-energy
effective action contains the terms which account for
the coupling to the background gravitational field:
$$
e^{S_{grav}} \sim e^{b(u)\chi + c(u) \sigma}
$$
Here $b(u), c(u)$ are the gravitational renormalization coefficients
computed for the low-energy $SU(2)$ theory in \cite{wittabl}.
We already know that
\be\label{know}
e^{2b} = {\Det} {{\p u^{k}}\over{\p a^{l}}} \ee
It remains to compute $e^{c}$.
Imagine that the manifold $X$ is replaced by a blowup $\widetilde X$
at the point $P$. Geometrically it means that we glue a copy of
$\bar\IP^{2}$ to $X$. The Euler characteristics and the signature
of the manifold $\widetilde X$ are $\chi + 1$ and $\sigma - 1$
respectively since $b_{2}^{-}$ is increased by one in the
process ($\bar\IP^{2}$ contains a non-contractible two-sphere
with self-intersection $-1$). Consider an integral of
the the divergence of the ghost number current over the glued
$\bar\IP^{2}$ in the instanton sector with the total instanton number $k$.
The instanton charge splits as a sum
$k = k_{1} + k_{2}$ where $k_{1}$ is what is left on $X - P$,
and $k_{2}$ is what has gone to $\bar\IP^{2}$. The divergence  picks
up an anomaly:
\be \label{anmly} \int_{\bar\IP^{2}} dJ = 2 \beta_{1} k_{2} \ee
where $\beta_{1}$ is the perturbative beta-function. In the case of pure
gauge theory it is equal to $-2h^{\vee}$, $h^{\vee}$ being  the dual Coxeter
number. There also could be a gravitational contribution to (\ref{anmly})
but it vanishes because $b_{2}^{+}({\bar \IP^{2}}) = 0$.
Now, using factorization we may replace the glued
$\bar \IP^{2}$
by a local operator $\CB (u^{k})$ which must produce  the same anomaly
as in (\ref{anmly}). If the theory is asymptotically free,
then the right hand side of (\ref{anmly}) is negative for $k_{2} > 0$
and therefore the operator $\CB = 1$ and only $k_{2} = 0$ contributes.
The conclusion is that the blowup does not change the measure.
On the other hand the gravitational renormalization together with
extra piece of Maxwell partition function combine to
\be\label{cnd}{e^{b - c} \Theta ({\tau}) = 1}\ee
where $\Theta (\tau)$ is a certain theta-constant which is
discussed below.
Hence:
\be\label{cndii}{e^{S_{grav}} \sim
{\rm Det}_{ij} \left( {{\p u^{i}}\over{\p a^{j}}} \right)^{{\chi +
\sigma}\over{2}} \Theta({\tau})^{\sigma}}\ee
It remains to notice that
$$
{\rm Det}_{ij} \left( {{\p u^{i}}\over{\p a^{j}}} \right)^{4}
\Theta^{8} = \Delta (u_{i})
$$
is the modular invariant function on the moduli space of
vacua (the discriminant) and
rewrite (\ref{cndii}) as
\be\label{cndiii}{e^{S_{grav}} \sim
{\rm Det}_{ij} \left( {{\p u^{i}}\over{\p a^{j}}} \right)^{{\chi}\over{2}}
\Delta^{{\sigma}\over{8}}}\ee

There is another gravitational correction to the effective action,
described in \cite{wittabl}, namely, if the manifold $X$ is not
spin, then there is a term $e^{{\half}(w_{2}(X), F)}$ in the
effective measure.
We can get rid of it in the course of the study of $SU(2)$ theory by
the shift $\tau \to \tau +1$ thanks to Wu formula.  The sign generalizes to
\be\label{curi}
e^{\left(\langle F, \rho \rangle, w_{2}(X)\right) }
\ee
in the case of general simply-laced group.
Suppose we are to integrate the massive $W$-bosons
out. Let $a \in \liet$ be the scalar in the Cartan part of the
vector multiplet. The one-loop determinants on bosons and fermions
cancel on non-zero modes leaving purely
holomorphic contribution:
\be\label{prtb}{\prod_{\alpha}
\langle a, \alpha \rangle^{\rm Ind_{d \oplus d^{*}}}
(L_{\alpha})}\ee
where $\alpha$ runs over the set of roots, $L_{\alpha}$ is the
line bundle corresponding to the root $\alpha$,
$\rm Ind_{d \oplus d^{*}}(L_{\alpha})$
is the index of the operator $d \oplus d^{*}$ coupled
to the line bundle $L_{\alpha}$ (the $W_{\alpha}$-boson multiplet).
The index formula gives ${{\rm Ind}_{d \oplus d^{*}}(L_{\alpha}) =
\half \left( \int_{X} c_{1}(L_{\alpha})^{2} +
c_{1}({X}) c_{1}(L_{\alpha}) \right) + {{\chi +\sigma}\over{4}}}$.
We see
that the terms $c_{1}(L_{\alpha})^{2}$ collect into the perturbative
beta-function, the term proportional
to ${{\chi + \sigma}\over{4
}}$ yields the asymptotics of
(\ref{cndiii}) in the $\Lambda \to 0$ limit and the odd part
$c_{1}(\Sigma) c_{1}(L_{\alpha})$ survives in the form
$\prod_{\alpha > 0} (-1)^{c_{1}({X}) c_{1}(L_{\alpha})}
\equiv    e^{\left(\langle F, \rho \rangle, w_{2}({X})\right) }$
since $c_{1}(X) \equiv w_{2}({X}) {\rm mod} 2$.
We denoted by $c_{1}(X)$ the first Chern class of the canonical bundle of the
almost complex structure which exists on any four-manifold
(cf. \cite{wittabl}).

\subsection{Specific computations of the contact terms}

\subsubsection{Two-observables}

Suppose we are interested in computing $\langle
\exp \int_{C_{a}} \CO^{(2)}_{\CP_{a}} \rangle_{X}$,
where $C_{a}$ are two-cycles on a four-manifold $X$ which may intersect.
If we go ahead and write down the effective low-energy measure we
immediately face the problem of modular anomaly. Indeed, the
low-energy measure contains a theta-function $\Theta_{H_{2}(X;{\IZ})}
({\tau})$\footnote{in the harmonic approach it becomes Siegel
theta-function associated to a lattice of signature $(b_{2}^{+}, b_{2}^{-})$}
associated to an intersection
form of $X$. It comes from the partition function of the Maxwell theory.
In the presence of the two-observables this
theta-function has an argument: $\Theta (\vec z, \tau)$,
$z_{i} = C_{a}^{\vee} {{\p \CP_{a}}\over{\p a^{i}}}$
and its modular transformation produces
a factor
\be\label{mdlan}
\exp \sum_{a,b} \# ( C_{a} \cap C_{b} )  ( \tau^{-1} )^{ij}
{{\p \CP_{a}}\over{\p a^{i}}} {{\p \CP_{b}}\over{\p a^{j}}}\ee
This factor must be cancelled by an anomaly of
an additional interaction which is developed due to the presence
of intersecting densities, i.e. due to contact terms.

In order to derive this interaction we again use the trick with
blow up of
a
manifold $X$ this time at the intersection point $P$.
The homology lattice  $H_{*}(\widetilde X)$
of $\widetilde X$ is that of $X$ plus a factor of
$\IZ$. The intersection form is simply
$$
(,)_{\widetilde X} = (,)_{X} \oplus (-1)
$$
as the exceptional divisor $e$ (the two-sphere inside
$\bar \IP^{2}$)
has self-intersection $-1$.
Under the isomorphism $H_{*}(\widetilde X) = H_{*}(X)
\oplus {\IZ}$
the inverse images of the cycles in $X$ belong to the component
$H_{*}(X)$ of $H_{*}(\widetilde X)$. We shall denote them
by the same letters as the cycles in $X$.
To derive the contact term we compare the correlation functions
$\langle \int_{C_{1}} \CO^{(2)}_{\CP_{1}} \int_{C_{2}} \CO^{(2)}_{\CP_{2}}
\ldots \rangle_{X}$
and
$\langle \int_{\tilde C_{1}} \CO^{(2)}_{\CP_{1}}
\int_{\tilde C_{2}} \CO^{(2)}_{\CP_{2}}
\ldots\rangle_{\widetilde X}$
where the cycles $\tilde C_{1}, \tilde C_{2} \in H_{*}({\widetilde X})$
do not intersect each other in the vicinity
of $P$ and are given by the formulae:
\be\label{lft}{{\tilde C}_{k} = C_{k} - e,\quad
 \# {\tilde C}_{1}\cap {\tilde C}_{2} =
\# C_{1}\cap C_{2} - 1}\ee
Consider the same ghost number current integral as in (\ref{anmly}).
Due to the presence of two $2$-observables $\int_{e} \CO^{(2)}_{\CP_{1,2}}$
the anomaly changes to:
\be\label{anmlyi} \int_{\bar\IP^{2}} dJ = 2\beta_{1} k_{2} + U(\CP_{1}) +
U(\CP_{2}) - 4 \ee
where $U(\ldots)$ denotes the ghost charge which equals twice the degree
of
$\ldots$ for homogeneous $\CP_{1,2}$. For simply-laced group and
for $k_{2} > 0$ the right hand
side of (\ref{anmlyi}) does not exceed $2 ( 2h - 2) - 2( 2h - T_{m}) =
2T_{m} - 4$, where $T_{m}$ is the contribution of matter  to the
perturbative
beta-function. Assuming that $T_{m} < 2$ we see that again the operator
$\CB$ must be equal to one (see \cite{frmor}, \cite{fintstern} for
mathematical proof of this result).
The net effect of our
manipulations is the replacement of the intersecting cycles on
the manifold $X$ by the non-intersecting cycles on the manifold
$\tilde X$.
Physically the crucial fact is that under the blowup of
a point $P$ a new two-cycle
$e$
appears and it leads to the possibility for the gauge field
to have a flux through it. In the low-energy effective theory
the insertion of the new two-cycle must be reflected in the
new factor in the Maxwell partition function, which is
the sum over all line bundles on $\overline{\IP}^{2}$
in the presence of two $2$-observables $\int_{e} \CO^{(2)}_{\CP_{1,2}}$
This new factor is simply:
\be\label{cnttrmii}
{{\p \CP_{1}}\over{\p a^{i}}}
{{\p \CP_{2}}\over{\p a^{j}}} {{\p}\over{\p \tau_{ij}}}
{\rm log} \Theta(\tau)
\ee
where
\be\label{tht}{\Theta = \sum_{\lambda \in \Lambda} \exp \left( 2\pi i \langle
\lambda, \tau \lambda\rangle  + \pi i \langle \lambda, \rho \rangle \right)
}
\ee
with $\Lambda$ being the set of weights and $\langle, \rangle$ the
restriction of the Killing form on $\liet$ - the Cartan subalgebra of $\lieg$.
The term $(-)^{\langle \lambda , \rho\rangle}$ is  (\ref{curi}) specified
to
$\bar \IP^{2}$ case.
The numerator $\p_{\tau} \Theta$ comes from evaluating the sum over
fluxes in the presence of two-observables while the denominator
$\Theta^{-1}$ is the remnant of the gravitational renormalization
factors $e^{b-c}$ from (\ref{cnd}).
Summarizing, we have shown that in the theories with $T_{m} < 2$
the pair contact term  is equal to (and more generally if
$U({\CP}_{1}) + U({\CP}_{2}) - 4 + 2\beta_{1} < 0$):
\be\label{prcnt}
{\CC}({\CP}_{1}, {\CP}_{2}) =
{{\p \CP_{1}}\over{\p a^{i}}}
{{\p \CP_{2}}\over{\p a^{j}}} {{\p}\over{\p \tau_{ij}}}
{\rm log} \Theta(\tau)
\ee

\subsection{Contact term  of $0$-observables and $4$-observables.}

Another way of getting the contact term $\CC( \CP_{1}, \CP_{2})$ is
by considering $4$- and $0$-observables. Let us consider asymptotically
free
theory
with massless matter.
Let $\CP_{1} = u_{1}$ be the generator whose fourth decsendant produces
the instanton charge (quadratic casimir). The insertion of
$e^{t\int_{X} u_{1}^{(4)}}$ is equivalent to multiplication of the
correlator in the sector with instanton charge $k$ by $e^{2\pi i t k}$.
It is possible to show that this manipulation is in turn equivalent to
rescaling of the fields $u_{k} \to e^{{{2\pi i d_{k}}\over{\beta_{1}}} t } u_{k}$
and $a^{i} \to e^{{4\pi i t}\over{\beta_{1}}} a^{i}$, where $d_{k}$ are the
weights
of the homogeneous generators $u_{k}$ of the ring of invariant polynomials.
This idea yields the following formula \cite{issues}:
\be\label{fz}
\CC ( u_{1}, u_{k} ) = {{4\pi i}\over{\beta_{1}}} \left(
a^{i} {{\p u_{k}}\over{\p a^{i}}} - d_{k}u_{k}\right) \ee

Comparing (\ref{fz}) and (\ref{prcnt}) in cases where it is possible
we get an interesting identity, which singles out the family of
SW curves and their generalizations \cite{SeWi},
\cite{argfar},\cite{argsei}. For concreteness we consider $SU(N_{c})$
theory with fundamental massless matter:
\be\label{idnty}
{{\p u_{1}}\over{\p a^{i}}}
{{\p u_{k}}\over{\p a^{j}}} {{\p}\over{\p \tau_{ij}}}
{\rm log} \Theta(\tau) =
{1\over{2N_c - N_{f}}}
\left( a^{i}
{{\p u_{k}}\over{\p a^{i}}} - (k+1) u_{k} \right) \ee
where$N_{f}$ is the number of
matter hypermultiplets $a^{i}$ are
the $A$-periods of the differential $x {{dz}\over{z}}$
on the curve:
$$
z + {{x^{N_{f}}}\over{z}} = x^{N_{c}} - \sum_{k=1}^{N_{c}-1} u_{k} x^{N_{c} - k -1}
$$
The formula (\ref{idnty}) must be valid for $N_{f} < N_{c}$ as follows
from the arguments under the formula (\ref{anmlyi}) applied to the
case $\CP_{1} = u_{1}$, $\CP_{2} = u_{k}$ as in this case
$U({\CP}_{1}) + U({\CP}_{2}) - 4 = 2k + 2 \leq 2N_{c}$
while $2\beta_{1} k_{2} \leq 2 (N_{f} - 2N_{c}) < - 2N_{c}$.

The formula (\ref{idnty}) follows from the modular properties
and the asymptotics at $u \to \infty$: both left and right hand sides
of (\ref{idnty})
vanish in the limit $u \to \infty$. The modular properties of the
left hand side follow easily from the theta constant behavior
while the right hand side enjoys the following property:
$$
{{2\pi i}\over{2N_c - N_{f}}}
\left( a^{i}
{{\p u_{k}}\over{\p a^{i}}} - (k+1) u_{k} \right) \vert_{D} -
{{2\pi i}\over{2N_c - N_{f}}}
\left( a^{i}
{{\p u_{k}}\over{\p a^{i}}} - (k+1) u_{k} \right) =
$$
$$
= {{2\pi i}\over{2N_{c} - N_{f}}}
\left( a_{D, j} - a^{i}\tau_{ij} \right) {{\p u_{k}}\over{\p a^{i}}}
\left( \tau^{-1} \right)^{ij} =
$$
$$
= {\half} {{\p u_{k}}\over{\p a^{i}}} {{\p u_{1}}\over{\p a^{j}}}
\left( \tau^{-1} \right)^{ij}
$$
thanks to the formulae of \cite{matone}.

It seems possible to
promote the formulae for the contact terms (\ref{prcnt})
and (\ref{fz}) to the {\it evolution} equations a la Whitham
hierarchy (cf. \cite{gmmm}, \cite{krichever}). Details will be shown
in a separate publication \cite{promise}.

\section{Remarks on ADHM integration}

Consider the moduli space $\CM_{k,N}$
of charge $k$ $U(N)$ instantons on $\IR^{4}$. By ADHM
construction it is the hyperkahler quotient of linear space
of dimension $4k^{2} + 4kN$ by the action of $U(k)$.
That is, it is the quotient of the space of solutions of certain
equations $\vec \mu = 0$ by $U(k)$. This representation allows
to get the expressions for the integrals of cohomology classes
over $\CM_{k,N}$  in terms of the contour integrals over the
complexified Lie algebra of $U(k)$.
The integrals are localized according to the standard
equivariant techniques \cite{atbott} to the fixed points of $U(k)$ action
which are nothing but the point-like instantons! The space of
those is non-compact - the instanton can run away to infinity.
In order to cure this problem let us take into account the natural
actions of the Lorentz group $SO(4)$ and gauge group $U(N)$.
Suppose we compute the integrals over $\CM_{k,N}$ of
the $SO(4) \times U(N)$-equivariant cohomology classes
$\omega_{1}(\vec \epsilon; \vec a), \ldots, \omega_{p}
(\vec \epsilon; \vec a)$.
Here $\vec \epsilon = {\epsilon_{1} \choose \epsilon_{2}}$ is the
generator of the Cartan subalgebra of $SO(4)$
and $\vec a = ( a_{1}, \ldots , a_{N})$ is the generator of $U(N)$
Cartan subalgebra. $\vec \epsilon$ generates rotations
in two orthogonal planes in $\IR^{4}$.
The fixed points
of such rotation are formed by the instantons which sit
on top of each other in the center of rotation.
Now the integral
can be evaluated (cf. \cite{vlm}). The classes $\omega_{l}$ are in one-to-one
correspondence with the invariant polynomials $P_{l}$ on the Lie algebra of
$U(k)$, i.e. symmetric functions in $k$ variables.
We claim that
\begin{eqnarray}
 Z ( \vec \epsilon ; \vec a) :=
\int_{\CM_{k,N}} \prod_{l} \omega_{l}(\vec \epsilon; \vec a) = & \nonumber\\
{{(\epsilon_{1} +\epsilon_{2})^{k}}\over{\epsilon_{1}^{k}\epsilon_{2}^{k}}}
\oint \bigwedge_{i=1}^{k}{{dz_{i}}\over{\prod_{\lambda} ( z_{i} +
a_{\lambda})
(z_{i} + a_{\lambda} + \epsilon_{1} + \epsilon_{2})}}
&  \prod_{i \neq j}
{{z_{ij} ( z_{ij} +\epsilon_{1} +\epsilon_{2})}\over{(z_{ij} +
\epsilon_{1})(z_{ij} + \epsilon_{2})}} \label{claim}
\end{eqnarray}
where $z_{ij} = z_{i} - z_{j}$.

The formula (\ref{claim}) is the only sensible calculation one may perform
using ADHM data. To compare it with the calculations involving
SW low-energy effective action one needs to develop a formalism
which takes into account the isometries of the space-time manifold.
It is equivalent
to working with $Q + \epsilon^{\mu}G_{\mu}$ - cohomology instead of
$Q$-cohomology (cf. \cite{los}, \cite{givental}).
This theory is currently under study \cite{promise}.

\acknowledgements

We thank G.~Moore for many discussions on Donaldson theory
and related matters over the years. S.Sh. is grateful to E.~Witten
for important discussion reviving the interest in the subject.
We are grateful to A.~Gerasimov, A.Gorsky, A.~Lawrence,
A.~Marshakov, A.~ Morozov, A.~ Mironov,
N.~Seiberg and C.~Vafa
for useful
discussions.

The research of A.~L.~ is supported partially by DOE under grant
DE-FG02-92ER40704, by
PYI grant PHY-9058501 and RFFI under grant 96-01-01101.
The research of N.~N.~ is supported by Harvard Society of Fellows,
partially by NSF under  grant
PHY-92-18167,  and partially by RFFI grant 96-02-18046.
The research of S.Sh. was supported by DOE grant DE-FG02-92ER40704,
by NSF CAREER award, by OJI award from DOE and by A.P.~Sloan
Foundation.
A.~L.~ and N.~N.~ are also supported by
grant 96-15-96455 for scientific schools.

\end{document}